\documentstyle[aps,eqsecnum,psfig]{revtex}
\newfont{\myfont}{cmmib10}

\newcommand{\half}{{\textstyle{\frac{1}{2}}}}

\newcommand{\brho}{\hbox{\myfont \symbol{26}} }

\def\mnras{Mon. Not. R. Astron. Soc.}
\def\araa{Ann. Rev. Astron. Astrophys.}
\begin{document}

\title{Circular Polarization Induced by Scintillation in a Magnetized
Medium}

\author{J.-P. Macquart$^{1}$ and D.B. Melrose$^{2}$}
\address{Research Centre for Theoretical Astrophysics\\
School of Physics, University of Sydney, NSW 2006, Australia.\\
$^{1}$jpm@physics.usyd.edu.au,
$^{2}$melrose@physics.usyd.edu.au
}

\date{\today}
\maketitle

\begin{abstract}
A new theory is presented for the development of circular polarization 
as radio waves propagate through the turbulent, birefringent 
interstellar medium.  The fourth order moments of the wavefield are 
calculated and it is shown that unpolarized incident radiation 
develops a nonzero variance in circular polarization.  A magnetized 
turbulent medium causes the Stokes parameters to
scintillate in a non-identical manner.  A specific model for this effect 
is developed for the case of density fluctuations in a uniform magnetic field.
\end{abstract}
\pacs{pacs \{98.58.Ay\} }
\pacs{ {\tt$\backslash$\string pacs\{98.58.Ay\}} }

\keywords{Faraday rotation --- polarization --- turbulence ---
pulsars: general --- galaxies: magnetic fields}

\section{Introduction}
Circular polarization observed in the radio emission from pulsars and 
quasars is not understood.  Although the emission mechanisms for 
these two classes of source are quite different, both have the common 
feature that the emission is due to highly relativistic particles in a 
magnetic field, for which the polarization should be predominantly 
linear with a circular component of order the inverse of the Lorentz 
factor of the radiating particles.  As discussed further below, this 
intrinsic component of circular polarization does not account for the 
observations.  We propose that the circular polarization is imposed 
as a propagation effect, due the scintillations in the interstellar 
medium \cite{Rickett77} which account well for many of the observed, 
and otherwise unexplained, time variations in pulsars and quasars 
\cite{Scheuer,Rickett84}.  The underlying physics for this alternative 
explanation is presented in this paper.  We propose to discuss the 
details of the astrophysical applications elsewhere.  

Scintillations are
attributed to scattering of the radio waves off density inhomogeneities
associated with turbulence in the ISM. The data indicate a
power-law model for the turbulence, with the power-law index
consistent with the Kolmogorov value ($\beta=11/3$ in the notation used
here) \cite{ArmRicSpa}. The dataset on pulsars is sufficiently large to allow
mapping of the turbulence across the Galaxy \cite{TaylorCordes}, implying much stronger
scattering at low Galactic latitudes, where most pulsars are observed,
than at high Galactic latitudes, where most quasars are observed.  A 
planar wavefront becomes rippled as it
traverses the region of turbulence.  Two length scales play a central 
role in the theory: the Fresnel scale,
$r_{\rm F}=(\lambda D/2\pi)^{1/2}$, where $\lambda$ is the wavelength of
the radio wave and $D$ is the distance between the observer and the
scattering region, and the diffractive scale, $r_{\rm diff}$, over which
the fluctuations in the phase decorrelate due to the turbulence. 
Physically, $r_{\rm diff}$ characterizes the
sizes of the ripples, and $r_{\rm F}$ characterizes the size of 
the coherent patch an observer can see on the unrippled wavefront when only the geometric
phase difference is taken into account.  ``Weak scattering'' 
corresponds to $r_{\rm F}\ll
r_{\rm diff}$, when an observer sees a single coherent patch that is slightly
tilted (image displacement) and slightly convex (focussed) or concave
(defocussed). ``Strong scattering'' corresponds to
$r_{\rm F}\gg r_{\rm diff}$, when an observer sees many coherent patches of
size $r_{\rm diff}$ within an envelope of size $r_{\rm ref}=
r_{\rm F}^2/r_{\rm diff}$, and multipath propagation occurs 
\cite{Narayan}.  Intensity variations in strong scattering are induced by both 
diffractive effects, due to interference between the coherent patches, 
and refractive effects, caused by focussing (defocussing) of the 
ray bundle due to phase curvature across the scattering disk.
There is a transition between strong scattering at lower frequencies and
weak scattering at higher frequencies. For most pulsars the transition
frequency is higher than the frequencies for which data are available,
and for quasars the observations span the expected transition frequency,
$\sim7\rm\,GHz$ \cite{Walker98}.

Inclusion of the interstellar magnetic field implies that the  ISM is
birefringent, and propagation of radiation depends upon its polarization.
In a homogeneous birefringent medium, radiation separates into the two
oppositely polarized natural wave modes, with the phase difference between
them increasing linearly with propagation distance. The two wavefronts
corresponding to the two modes become systematically displaced from each
other with increasing distance. The natural modes in the ISM are
circularly polarized to an excellent approximation, and the birefringence
results in Faraday rotation of the plane of linear polarization of any
incident radiation. The amount of Faraday rotation is parametrized 
in terms of the rotation measure, RM, which is
defined such that the phase difference between the two modes is
RM$\lambda^2$ \cite{Ratcliffe59}. Inhomogeneity combined with birefringence implies that the
wavefronts are both rippled and displaced from each other. One
implication is that when the wavefronts are recombined there is a random
component in the phase difference, and the resulting stochastic Faraday
rotation is characterized by a variance in RM \cite{MelMac}.

The main point made in the present paper is that scattering in the 
magnetized ISM necessarily leads to a component of circular 
polarization (CP) in the observed radiation.  This arises because any 
lateral displacement of the wavefront implies that the ripples are not
superimposed when the two wavefronts are combined.  As a result, there
are alternative patches of excess right hand (RH) and excess left hand
(LH) circular polarization on the image in the observer's plane.  In 
this paper we present a detailed theory for such 
scintillation-induced CP.  We argue that the predicted features of 
the CP are sufficiently promising to warrant the development of 
detailed models for the observed CP in pulsars and quasars based on it.

CP is observed in both pulsars \cite{Manchester71} and in some compact extragalactic
sources \cite{JARob,Weiler,dePater,Kom}. Most of the data on pulsars are integrated over many
pulses, and the integrated pulse profile typically shows relatively small
CP. However, there is evidence that in at least a few pulsars for which
individual pulses can be studied, the CP is relatively large in
individual pulses and varies from pulse to pulse such that the integrated
value is much smaller than the typical value. There is no satisfactory
explanation for the CP \cite{radhakrishnan}.  A small ($\lesssim 0.1$\% to
a few
\%) but significant  degree of CP is observed in some compact 
extragalactic sources
\cite{JARob,Weiler,dePater,Kom}. The suggested interpretations include
the intrinsic polarization associated with  synchrotron radiation
\cite{LeggWest}, and  partial conversion of linear into
circular polarization due to ellipticity of the natural wave modes of
the cold background plasma \cite{Pachol} or of the relativistic electron gas
itself \cite{Sazonov,Jones77a,Jones77b}.  However none of these
suggested interpretations has proved satisfactory in accounting for
(a) the frequency dependence, (b) the temporal variations, and (c)
the magnitude of the observed circular polarization \cite{saikia}.
The explanation proposed in this paper is as outlined above. Specifically,
for a source with zero intrinsic CP seen through a turbulent magnetized
plasma, there is a variable CP which corresponds to a zero average of the
Stokes parameter  $V$, $\langle V \rangle=0$, but a nonzero variance,
$\langle V^2 \rangle \neq 0$. Our initial objective is to use the theory
of scintillations in a magnetized plasma
\cite{Eru,Tamoikin,Melrose93a,Melrose93b} to calculate $\langle
V^2\rangle$.

The magnitude of the expected value of CP needs to be of the
same order of magnitude as the observed CP for the theory
proposed here to be relevant. For pulsars, the CP can be several
tens of percent, but it may be that some of this CP results from
birefringence in the pulsar magnetosphere itself, which we do not
consider in detail here. For the most extreme case for quasars, the
observed CP can be several percent, which is relatively high
because there is independent evidence suggesting that the varying
(scintillating in our interpretation) part of the source is only a
modest fraction of the entire source. Hence, the observations, in
the most extreme cases, suggest that the CP of the scintillating
component can be as high as a few tens of percent. In the theory
developed here, most of the terms that contribute to the CP are very
small, of order the ratio of the cyclotron frequency in the ISM to
the radio frequency (typically $\sim10^{-8}$), and can be of no
practical interest. However, the effect on which we concentrate can
give arbitrarily high CP. This effect is due to birefringent
refraction causing an angular separation between the emerging rays
in the LH and RH polarized wave modes. The displacement of the
centroids of the LH and RH images increases linearly with the
distance from the screen where the birefringent refraction occurs.
It is possible for the LH and RH images not to overlap, resulting in
patches of 100\% CP. The angular deviation required to produce
relatively large CP is determined by the ratio of the characteristic
size of the scintillation pattern divided by the distance between
the observer and the screen where the birefringent refraction
occurs. Although the angular separation between the rays is always
extremely small, we argue elsewhere\cite{MMcp2} that observed
gradients in RM imply birefringent refraction through a sufficiently
large angle to satisfy the criterion that observable CP be produced.

Our specific assumptions are explained in \S2, where the wave equation
is reduced to the form used in treating the scattering.   The
mutual coherence of a polarized wavefield is derived  in \S3, and is
shown to reproduce some known results \cite{Kuk91a,Kuk91b}.  In \S4 the
second-order correlations of the  Stokes parameters are derived from the
fourth-order moment of  the wavefield, and explicit solutions are
obtained in the thin-screen approximation.  In \S5 we  discuss
scintillation-induced CP. The conclusions are presented in \S6.

\section{Propagation through a magnetized plasma}
In this section the propagation of radiation through a magnetized stochastic
medium is related to its
effect upon the electric field of the wave.  We start with propagation
through a homogeneous weakly anisotropic medium and then generalize to
include the effect of inhomogeneities in the scattering medium.  For a
weakly inhomogeneous medium the two wave modes are assumed to be
transverse to a zeroth approximation (the isotropic limit) and
the degeneracy between the two transverse
states of polarization is broken by the weak
anisotropy, to a first approximation.  In the first approximation the two
modes have slightly different refractive
indices, and this approximation suffices to treat Faraday rotation
and all the effects of interest here.

The wave equation projected onto the transverse plane is
\cite{Melrose93a}
\begin{eqnarray}
\left(-k^2 \delta^{\alpha \beta} +
\frac{\omega^2}{c^2} K^{\alpha \beta}(\omega,{\bf k}) \right)
A^{\beta}(\omega,{\bf k})= 0,    \label{WaveEqn}
\end{eqnarray}
where $A^{\alpha}(\omega,{\bf k})$ is the wave amplitude,
$K^{\alpha \beta}(\omega,{\bf k})$ is the
dielectric tensor and the greek
indices run over the two transverse co-ordinates.  We write
\begin{eqnarray}
    K^{\alpha \beta} = \langle K^{\alpha \beta} \rangle +
    \delta
    K^{\alpha \beta},
\end{eqnarray}
where the angular brackets denote the mean and $\delta K^{\alpha
\beta}$ denotes a fluctuating part with a mean of zero, $\langle \delta
K^{\alpha \beta} \rangle=0$.  The
transverse components of the dielectric tensor may be expressed in
terms of the Pauli matrices, and for the average part we write
\begin{eqnarray}
\langle K^{\alpha\beta} \rangle=K_{A}\sigma_A^{\alpha\beta},
\end{eqnarray}
where the sum over $A=[I,Q,U,V]$ is implied, with
\begin{eqnarray}
\sigma_I^{\alpha\beta}=\delta^{\alpha\beta}
=\pmatrix{1&0\cr0&1\cr},
\quad
\sigma_Q^{\alpha\beta}=\left(
\begin{array}{rr} 1&\!\!0 \\0&\!\!-1
\end{array} \right),
\quad
\sigma_U^{\alpha\beta}=\pmatrix{0&1\cr1&0\cr},
\quad
\sigma_V^{\alpha\beta}=\left( \begin{array}{rr} 0&\!\!-i \\
i&\!\!0 \end{array}
\right).
\end{eqnarray}
In the discussion here we neglect any dissipation, which implies that
we retain only the Hermitian part of $K^{\alpha \beta}$, so the $K_A$
are all real.

The two modes are labelled $\sigma=\pm$.  The mode $\sigma$ has
wavenumber $k_{\sigma}=n_{\sigma}
\omega/c$, where $n_{\sigma}$ is the refractive index for waves in
mode $\sigma$, and polarization
vector ${\bf e}_{\sigma}$.  It is straightforward to solve for the
dispersion relations, which are
\begin{eqnarray}
k^2_\sigma={\omega^2\over c^2}
\big\{K_I+\sigma
\big[K_Q^2+K_U^2+K_V^2\big]^{1/2}\big\}.
\end{eqnarray}
The quantity $K_I$ is the dielectric constant in the isotropic
approximation, with $K_Q^2+K_U^2 +K_V^2 \ll K_I^2$ in the weak anisotropy
limit.
The polarization vectors in the two-dimensional transverse plane are
\begin{eqnarray}
{\bf e}_\sigma=
{\big(K_Q+\sigma
\big[K_Q^2+K_U^2+K_V^2\big]^{1/2},
K_U+iK_V
\big)
\over\big\{
2\big[K_Q^2+K_U^2+K_V^2\big]^{1/2}
\big(K_Q+\sigma
\big[K_Q^2+K_U^2+K_V^2\big]^{1/2}\big)
\big\}^{1/2}}. \label{esigma}
\end{eqnarray}
One is always free to orient the coordinate axes such that
$K_U=0$.  On doing so, equation (\ref{esigma}) may be
written in the form
\begin{eqnarray}
{\bf e}_\sigma=(1,iT_\sigma),
\qquad
T_\sigma={K_V\over K_Q+\sigma
\big[K_Q^2+K_V^2\big]^{1/2}},
\qquad
T_+T_-=-1,
\end{eqnarray}
where $T_\sigma$ is the axial ratio of the polarization ellipse
in the mode $\sigma$.
In the case of Faraday rotation, the wave modes are circularly polarized,
which corresponds to $| T_{\sigma} | = 1$.  The circular polarizations
are
\begin{eqnarray}
{\bf e}_{r,l} = \frac{1}{\sqrt{2}} (1,\pm i),
\end{eqnarray}
where $r$ and $l$ refer to right- and left-hand respectively.
In this paper we consider only the circularly polarized approximation
explicitly, but the general theory is valid for any $T_{\sigma}$.

In scattering theory it is conventional to make the parabolic approximation
to the wave equation \cite{Tatarskii,Ishimaru}.  This
approximation is related to the paraxial approximation in geometric
optics, in the sense that there is a favored ray direction (the
$z$-axis in our case) and that only small deviations from it are
considered.  In
the parabolic approximation the wavefield is written as the product of a
fast-varying term $e^{i k_\sigma z}$ and a term
$u^\alpha(z,{\bf r})$ that is assumed to be a slowly varying
function of $z$ in the sense that its second derivative with respect to $z$
may be neglected.  The difference between $k_+$ and $k_-$ is
introduced explicitly by writing  $k_\sigma= k + \sigma \delta k$,
which defines $\delta k$.  In a homogeneous medium the two phase
factors $ik_{\sigma} z$ include the mean phase $ikz$, corresponding to
the average over the two modes, and the phase difference 
$\pm i \, \delta k\, z$ between the components in the two modes relative to this
mean.

The inhomogeneities are introduced through a fluctuating part,
$\delta K^{\alpha \beta}$, of the dielectric tensor.  On making the
parabolic approximation to equation (\ref{WaveEqn}),
one obtains the following equations for the propagation of
the right- and left-hand polarized wavefields
\begin{eqnarray}
    \left( 2 i k \frac{\partial}{\partial z} + \nabla_\perp^2
    +\frac{\omega^2}{c^2}  \delta K_{++} \right)\tilde u_+ +
    \frac{\omega^2}{c^2}  \delta K_{+-} \tilde u_- e^{-2 i \delta k z}= 0,
\nonumber \\
    \left( 2 i k \frac{\partial}{\partial z} + \nabla_\perp^2
    +\frac{\omega^2}{c^2}  \delta K_{--} \right) \tilde u_- +
   \frac{\omega^2}{c^2}  \delta K_{-+} \tilde u_+ e^{2 i \delta k z}=
   0, \label{fulleqs}
\end{eqnarray}
where the perturbation terms involve
\begin{eqnarray}
\delta K_{\sigma\sigma'}=e^{*\alpha}_\sigma
e^\beta_{\sigma'}\delta K^{\alpha\beta},
\end{eqnarray}
and
\begin{eqnarray}
     A^\alpha = \sum_{\sigma=\pm} e^{i k z} {\bf e}_\sigma
    \tilde{\bf u}_\sigma (z,{\bf r}), \qquad \qquad \tilde{\bf u}_\sigma
(z,{\bf
    r})= {\bf e}_{\sigma} u_\sigma (z,{\bf r}) e^{i\sigma \delta k
    z}. \label{ParabolicApprox}
\end{eqnarray}

The terms involving $\delta K_{+-}$ and $\delta K_{-+}$ are zero if
the inhomogeneities do not affect the polarization of the natural
modes, that is if $\delta K^{\alpha \beta} - \sigma_I^{\alpha \beta}
\delta K_I$ is proportional to
$\langle K^{\alpha \beta} \rangle - \sigma_I^{\alpha \beta} \langle K_I
\rangle$.
This is the case to an excellent approximation if the fluctuations do
not involve the direction of the magnetic field, and even for
fluctuations that affect the direction of the magnetic field it is an
excellent approximation provided that the modes are nearly circularly
polarized.  The extreme conditions under which these terms might be
non-negligible are ignored here.

Neglecting the terms involving $\delta K_{+-}$ and $\delta K_{-+}$, the
following
relations describe the propagation of the wave amplitude through a
weakly anisotropic inhomogeneous medium
\begin{eqnarray}
    \left( 2 i k \frac{\partial}{\partial z} + \nabla_\perp^2
    +\frac{\omega^2}{c^2}  \delta K_{++} \right)\tilde u_+ = 0, \nonumber \\
    \left( 2 i k \frac{\partial}{\partial z} + \nabla_\perp^2
    +\frac{\omega^2}{c^2}  \delta K_{--} \right) \tilde u_-= 0.
\label{Uprop}
\end{eqnarray}
Equations (\ref{Uprop}) are used as the basis for the theory developed 
below.

\section{Second order moments of the wavefield}
In this
section we calculate the average visibilities in each of the four
Stokes parameters, thus determining the properties of the average
image of a scattered source.  

\subsection{Ensemble average visibilities}
Consider a two-element interferometer with receivers
located at positions ${\bf r}_1$ and ${\bf r}_2$, that 
measures the left- and right-hand circularly polarized
components of the electric field.  We define the visibility in a 
given polarization using the generalized second-order moment of 
the electric field
\begin{eqnarray}
\gamma_{\sigma \sigma'}(z;{\bf r}_1,{\bf r}_2) =
\langle u_{\sigma}(z;{\bf r}_1) u_{\sigma'}^*(z;{\bf r}_2) \rangle.
\end{eqnarray}
The Stokes parameters are
defined in terms of the left- and right-hand circularly polarized
components of the electric field for ${\bf r}_1={\bf r}_2$: 
$I_{++} = u_+ u_+^*$, $I_{--}=u_-
u_-^*$, $I_{+-} = u_+ u_-^*$ and $I_{-+} = u_- u_+^*$.  These are
related to the conventional Stokes parameters $I$, $Q$, $U$ and $V$
by
\begin{eqnarray}
I=\half(I_{++}+I_{--}),
\qquad
Q=\half(I_{+-}+I_{-+}),
\qquad
U=i\half(I_{+-}-I_{-+}),
\qquad
V=\half(I_{++}-I_{--}),
\hfill\cr
\noalign{\vskip3pt}\hfill
I_{++}=I+V,
\qquad
I_{--}=I-V,
\qquad
I_{+-}=Q-iU,
\qquad
I_{-+}=Q+iU.
\hfill
\label{StokesTwo}
\end{eqnarray}
In this notation the Stokes visibilities are
\begin{eqnarray}
I(z;{\bf r}_1,{\bf r}_2)&=&\half[\gamma_{++}(z;{\bf r}_1,{\bf r}_2) +
\gamma_{--}(z;{\bf
r}_1,{\bf r}_2)], \nonumber \\
Q(z;{\bf r}_1,{\bf r}_2)&=&\half[\gamma_{+-}(z;{\bf r}_1,{\bf r}_2) +
\gamma_{-+}(z;{\bf
r}_1,{\bf r}_2)], \nonumber \\
\hfill U(z;{\bf r}_1,{\bf r}_2)&=&i\half[\gamma_{+-}(z;{\bf r}_1,{\bf r}_2)
- \gamma_{-
+}(z;{\bf r}_1,{\bf r}_2)], \nonumber \\
V(z;{\bf r}_1,{\bf r}_2)&=&\half[\gamma_{++}(z;{\bf r}_1,{\bf r}_2) -
\gamma_{--}(z;{\bf
r}_1,{\bf r}_2)].
\label{StokesVis}
\end{eqnarray}
Using equations (\ref{Uprop}) and their complex
conjugates, the four visibilities obey the four propagation equations
obtained by setting $\sigma=\pm 1$ and $\sigma' = \pm 1$ in
\begin{eqnarray}
\left[ 2 i k \frac{\partial}{\partial z}+ \nabla_1^2 - \nabla_2^2 +
k^2 \delta K_{\sigma \sigma}({\bf r}_1) - k^2 \delta K_{\sigma' \sigma'}({\bf
r}_2) \right]
\gamma_{\sigma \sigma'}(z;{\bf r}_1,{\bf r}_2) e^{i \delta k z
(\sigma' - \sigma)}= 0. \label{unavgdGamma}
\end{eqnarray}
Due to both the properties of the radiation from the source and
the stochastic nature of the phase screen, it is desirable to
compute the ensemble average visibility, denoted by $\Gamma_{\sigma
\sigma'}$.  The ensemble average is considered an average over both
time and over the phase fluctuations.  It is normally
assumed that the phase screen itself is static, with any perceived
time-variability due to relative motion between the screen and the
source-observer line of sight at some velocity ${\bf v}$.  This is
known as the frozen screen approximation.  An observer measures the visibility
function at time $t$ to be $\gamma_{\sigma \sigma'}(z;{\bf r}_1+{\bf
v}t,{\bf r}_2+{\bf v}t)$.  However, if the system is homogeneous this
visibility depends only on ${\bf r}_1+{\bf v}t - {\bf r}_2 - {\bf v}t$ and
is therefore independent of $t$.  Thus the average over time is
trivial and the visibility is a function of relative receiver separation
only.

On replacing the independent variables ${\bf r}_1$ and
${\bf r}_2$ by ${\bf r} = {\bf r}_1-{\bf r}_2$,
${\bf s}=\half({\bf r}_1+{\bf r}_2)$, it follows that in a
statistically homogeneous medium the average over the fluctuations is a
function of ${\bf r}$ only.  The average over the propagation
equation (\ref{unavgdGamma}) then leads
to a propagation equation for, $\Gamma_{\sigma \sigma'} (z;{\bf r}) = \langle
\gamma_{\sigma \sigma'}(z;{\bf r}+{\bf r}',{\bf r}') \rangle$:
\begin{eqnarray}
\left[ 2 i k \frac{\partial}{\partial z}+ \nabla_{{\bf r}\cdot{\bf s}} +
k^2 \xi_{\sigma \sigma'}({\bf r}) \right] \Gamma_{\sigma
\sigma'}(z;{\bf r}) e^{i \delta k z (\sigma' - \sigma)} = 0, \\
\xi_{\sigma \sigma'}({\bf r}) = \langle \delta K_{\sigma \sigma}(0) -
\delta K_{\sigma' \sigma'}({\bf r}) \rangle,
\end{eqnarray}
with $\nabla_{{\bf r}\cdot {\bf s}} = 2 (\partial^{2}/\partial r_x \partial s_x -
\partial^{2}/\partial r_y \partial s_y) $, ${\bf r}=r_x \hat{\bf
x}+r_y \hat{\bf y}$, ${\bf s}=s_x \hat{\bf x}+s_y \hat{\bf y}$.
Suppose that the phase inhomogeneities are located on a thin
screen of thickness $\Delta z$.  Then to first order in $\Delta z$, the
visibility measured a distance $z$ from the screen is independent of the
screen thickness.  It is related to the incident
visibility $\Gamma_{\sigma \sigma'}(0;{\bf r}')$ according to
\begin{eqnarray}
\Gamma_{\sigma \sigma'}(z;{\bf r}) &=& \left(\frac{k}{2 \pi z}\right)^2 e^{i \delta
k (\sigma' -\sigma)}
\int d^2{\bf r}' \, d^2{\bf s}' \Gamma_{\sigma \sigma'}(0;{\bf r}') \,\exp \left[
\frac{ik({\bf r}-{\bf r}')\cdot {\bf s}'}{z} + i \Delta
\phi_{\sigma \sigma'} ({\bf r}')  \right], \\
\Delta \phi_{\sigma \sigma'}(z;{\bf r}') &=& k^2 \int_0^z dz'
\xi_{\sigma
\sigma'}(z';{\bf r}').
\end{eqnarray}

\subsection{The average over phase fluctuations}\label{AvgPhaseSec}

The average over the random fluctuations on the screen is
performed under the assumption that the fluctuations in both the isotropic and
anisotropic terms are gaussian.
The average $\langle \exp[i \delta x] \rangle = \exp[-\langle (\delta x)^2
\rangle/2]$ applies for any gaussian random variable $\delta x$.
Writing  $\delta \phi_{\sigma}({\bf r}) =
\int_0^z dz' (\omega/c) \delta K_{\sigma \sigma}(z';{\bf r})$, we make
use of this average in defining the generalized phase structure function:
\begin{eqnarray}
D_{\sigma \sigma'}({\bf r}) &=& 2 [C_{\sigma \sigma'}(0) -
C_{\sigma \sigma'}({\bf r})], \label{Ddefn} \\
C_{\sigma \sigma}({\bf r}) &=& \langle \delta \phi_{\sigma}({\bf r}')
\delta \phi_{\sigma'}({\bf r}'+{\bf r}) \rangle.
\end{eqnarray}
On specializing to the case where the natural wave modes are
circularly polarized, the contribution of
the isotropic and anisotropic fluctuations in equation
(\ref{Ddefn}) may be made explicit by
introducing the notation $\delta K_{\sigma \sigma} = \delta K_I +
\sigma K_{V}$, where $K_{I}$ is the isotropic component of the tensor
and $K_V$ is the anisotropic component.  (The ellipticity of the
modes is determined by $K_Q/K_V$, which is set to zero in assuming
that the polarizations are circular.)  The phase fluctuations may
be separated in an identical manner:
$\delta \phi_\sigma({\bf r}) = \delta \phi_I({\bf r}) +
\sigma \delta \phi_V({\bf r})$ with $\phi_I$ and $\phi_V$ denoting the
isotropic and anisotropic components, respectively.  Equations
(\ref{Ddefn}) are expanded as follows
\begin{eqnarray}
D_{\pm \pm}({\bf r}) &=& D_{II}({\bf r}) \pm 2 D_{IV}({\bf r})
+ D_{VV}({\bf r}) \\
D_{\pm \mp}({\bf r}) &=& D_{II}({\bf r}) \mp D_{IV}({\bf r}) \pm
D_{VI}({\bf r}) - D_{VV}({\bf r}) + 4 C_{VV}(0),
\end{eqnarray}
with
\begin{eqnarray}
C_{XY}  = \langle \delta \phi_{X}({\bf r})
\delta \phi_{Y}({\bf r}'+{\bf r}) \rangle, \qquad \qquad X,Y = [I,V].
\end{eqnarray}
The structure function $D_{II}({\bf r})$ represents the effect of the
isotropic phase fluctuations.  Fluctuations purely in the rotation
measure are characterized by $D_{VV}$ which we call the
``rotation measure structure function''.  The terms $D_{IV}$ and $D_{VI}$
represent the cross-correlations between in the isotropic and anisotropic
phase
fluctuations.

Performing the average over the phase fluctuations and
denoting the mean anisotropic phase $\delta k \,z/2$ introduced in equation
(\ref{ParabolicApprox}) by $-\phi_{V}$ the ensemble-average mutual coherence is
\begin{eqnarray}
\Gamma_{\sigma \sigma'}(z;{\bf r}) = \left(\frac{k}{2 \pi z}\right)^2 \int d^2{\bf r}' 
\, d^2{\bf s}'
\Gamma_{\sigma \sigma'}(0;{\bf r}')\, \exp \left[ \frac{ik({\bf r}-
{\bf r}')\cdot {\bf s}'}{z}-\frac{D_{\sigma \sigma'}({\bf r}')}{2} + i (\sigma -
\sigma') \phi_{V}
\right]. \label{2ptcoher}
\end{eqnarray}
Equation (\ref{StokesVis}) is inverted at $z=0$ to obtain
initial values of $\Gamma_{\sigma \sigma'}$ in terms of the Stokes
parameters at the screen $z=0$, $I(0;{\bf r})$,
$Q(0;{\bf r})$, $U(0;{\bf r})$ and $V(0;{\bf r})$.  The ensemble-average
visibilities may therefore be expressed in terms of the initial
polarization and the generalized phase structure function as follows
\begin{eqnarray}
   \langle I \rangle (z;{\bf r}) &=&
   \frac{ I(0;{\bf r})+V(0;{\bf r})}{2} \exp[-\half D_{++}({\bf r})] +
   \frac{I(0;{\bf r})- V(0;{\bf r})}{2} \exp[-
\half D_{--}({\bf r})],  \nonumber \\
\langle Q \rangle (z;{\bf r})  &=&
 \frac{Q(0;{\bf r})-iU(0;{\bf r})}{2} e^{2 i \phi_{V} }
\exp[-\half D_{+-}({\bf r})] + \frac{Q(0;{\bf r})+iU(0;{\bf r})}{2}
e^{-2 i \phi_{V} }
\exp[-\half D_{-+}({\bf r})],  \nonumber \\
\langle U \rangle (z;{\bf r}) &=& \frac{iQ(0;{\bf r})+U(0;{\bf r})}{2}
e^{2 i \phi_{V} } \exp[-\half D_{+-}({\bf
r})] - \frac{iQ(0;{\bf r})-U(0;{\bf r})}{2} e^{-2 i \phi_{V} }
\exp[-\half D_{-+}({\bf r})],
\nonumber \\
\langle V \rangle (z;{\bf r})  &=& \frac{I(0;{\bf r})+V(0;{\bf r})}{2}
\exp[-\half D_{++}({\bf r})] - \frac{I(0;{\bf r})-V(0;{\bf r})}{2}
\exp[-\half D_{--}({\bf r})].  \label{StokesVisFull}
\end{eqnarray}

The assumption that the statistics of the phase fluctuations are
stationary,  implies $\langle \delta \phi_I({\bf r}) \delta \phi_V({\bf
r}+{\bf r}') \rangle = \langle \delta \phi_V({\bf r})
\delta \phi_I({\bf r}+{\bf r}') \rangle$, and hence $D_{IV} = D_{VI}$. 
The visibilities reduce to 
\begin{eqnarray}
\left( \begin{array}{l}
\langle I \rangle (z;{\bf r}) \\
\langle Q \rangle (z;{\bf r}) \\
\langle U \rangle (z;{\bf r}) \\
\langle V \rangle (z;{\bf r}) \\
\end{array} \right) &=& e^{-D_{II}({\bf r})/2}  \left(
\begin{array}{c}
e^{- D_{VV}({\bf r})/2} [\langle I \rangle (0;{\bf r}) \cosh D_{IV}({\bf
r}) - \langle V \rangle (0;{\bf r}) \sinh D_{IV}({\bf r}) ] \\
e^{ D_{VV}({\bf r})/2 - 2 C_{VV}(0)}
 \{ \langle Q \rangle (0;{\bf r}) \cos 2 \phi_V + \langle U
\rangle (0;{\bf r}) \sin 2 \phi_V \} \\
e^{D_{VV}({\bf r})/2 - 2 C_{VV}(0)}
\{ - \langle Q \rangle (0;{\bf r}) \sin 2 \phi_V + \langle
U\rangle(0;{\bf r}) \cos 2
\phi_V \} \\
e^{- D_{VV}({\bf r})/2} [-\langle I \rangle (0;{\bf r}) \sinh
D_{IV}({\bf r}) + \langle V \rangle (0;{\bf r}) \cosh D_{IV}({\bf r}) ]
\end{array} \right). \label{MeanImg}
\end{eqnarray}
The mean values of the Stokes parameters $I$ and $V$ are equal to
their incident values; this may be seen by setting ${\bf r}$ equal
to zero in equation (\ref{StokesVisFull}),
and noting $D_{\sigma \sigma'}(0) = 0$ according to equation
(\ref{Ddefn}).  However, even if the initial circular polarization is
zero, its visibility, $\langle V \rangle (z;{\bf r})$
is nonetheless non-zero by virtue of the
difference between $D_{++}({\bf r})$ and $D_{--}({\bf r})$ at nonzero
${\bf r}$.  An interferometer may, as opposed to a single dish, therefore
detect
circular polarization from a source even if its radiation is
intrinsically unpolarized.  This effect was identified by
Kukushkin \& Ol'yak \cite{Kuk91a,Kuk91b}.

The depolarization of the linear Stokes visibilities due to stochastic
Faraday rotation is manifested through the term
$\exp[-2 C_{VV}(0)]$ \cite{MelMac}. 

In a simple model for rotation measure fluctuations in a homogeneous 
magnetic field (see \S \ref{CoherMag}), the 
resulting circularly polarized visibility, $\langle V \rangle (z;{\bf r})$, 
is of order $\alpha$ times smaller than the mean intensity for initially 
unpolarized radiation.  Effects of order $\alpha$ are far too 
small to be of interest for the ISM.
Nevertheless, it is of formal interest to interpret their origin.  In a
medium with a homogeneous magnetic field one has $\delta
\phi_V = \alpha \phi_I$, and the phase structure function for
the right-hand polarized wavefront is
$(1+\alpha)^2 D_{II}({\bf r})$ and that for the left-hand polarized wavefront
is $(1-\alpha)^2 D_{II}({\bf r})$.  Thus
the scale-length over which each wavefront experiences a
root-mean-square phase difference of one radian differs slightly.  This is
interpreted in terms of each
sense of circular polarization corresponding to different parameters,
${r_{\rm diff}}_+$ and ${r_{\rm diff}}_-$, say, for the diffractive
scales.  If $\alpha$ is positive, one has
${r_{\rm diff}}_+ < {r_{\rm diff}}_-$, and
the left-hand polarized visibilities extend to larger baselines
than the right-hand polarized visibilities.

\section{Fourth order moments of the wavefield}
The discussion in \S3 refers to the ensemble averages of the Stokes
visibilities.  In this section we derive the variance of the
visibilities and hence of the Stokes parameters themselves.
The underlying idea is that $\langle V^2 \rangle \neq 0$ and $\langle
V \rangle =0$ can lead to observable circular polarization provided
that the timescale for the fluctuations in $V$ is long compared with
the timescale for an observation.

\subsection{Solution for statistically homogeneous fluctuations}
The fourth order moment of the wavefield describes the correlations of
the Stokes visibilities.  Using the definitions of
the Stokes parameters in equation
(\ref{StokesTwo}) the auto- and cross-correlations of the Stokes
visibilities may be
expressed in terms of the following generalized fourth-order
moment of the electric field
\begin{eqnarray}
\gamma_{\sigma_1 \sigma_2 \sigma_1' \sigma_2'}(z;{\bf r}_1,{\bf
r}_2,{\bf r}_1',{\bf r}_2') =
\langle u_{\sigma_1}(z;{\bf r}_1) u_{\sigma_2}(z;{\bf r}_2)
u_{\sigma_1'}^*(z;{\bf r}_1') u_{\sigma_2'}^*(z;{\bf r}_2') \rangle.
\end{eqnarray}
This moment describes the
cross-correlation in the electric field between four receivers at
positions ${\bf r}_1$, ${\bf r}_1'$,
${\bf r}_2$ and ${\bf r}_2'$, each receiver measuring either the
right- or left-hand circularly polarized component of the radiation
according to the sign of the subscript $\sigma_1, \sigma_2,
\sigma_1', \sigma_2' = \pm 1$.  

Equations (\ref{Uprop}) are used to derive the following equation
for generalized fourth-order moment (actually 16 equations for the 16
moments):
\begin{eqnarray}
\left( 2 i k \frac{\partial}{\partial z} + \nabla_1^2 + \nabla_2^2 -
\nabla_1'^2 - \nabla_2'^2 + i k G_{\sigma_1 \sigma_2 \sigma_1'
\sigma_2'}' (z;{\bf r}_1,{\bf
r}_2,{\bf r}_1',{\bf r}_2') \right) \gamma_{\sigma_1 \sigma_2 \sigma_1'
\sigma_2'}(z;{\bf r}_1,{\bf
r}_2,{\bf r}_1',{\bf r}_2') = 0, \label{4-2}
\end{eqnarray}
with $G_{\sigma_1 \sigma_2 \sigma_1' \sigma_2'}'$ being the
ensemble average over the phase fluctuations.  For
$\sigma_1 + \sigma_2 - \sigma_1' - \sigma_2' \neq 0$ the
following result holds:
\begin{eqnarray}
G_{\sigma_1 \sigma_2 \sigma_1' \sigma_2'}'(z;{\bf r}_1,{\bf
r}_2,{\bf r}_1',{\bf r}_2') &=& D'_{z;\sigma_1 \sigma_1'}({\bf r}_1-
{\bf r}_1') + D'_{\sigma_1 \sigma_2'}(z;{\bf r}_1-{\bf r}_2') +
D'_{\sigma_2 \sigma_1'}(z;{\bf r}_2-{\bf r}_1') + D'_{\sigma_2
\sigma_2'}(z;{\bf r}_2-{\bf r}_2') \nonumber \\ &\null&
- D'_{\sigma_1 \sigma_2}(z;{\bf r}_1-{\bf r}_2) -
D'_{z;\sigma_1' \sigma_2'}({\bf r}_1'-{\bf r}_2')+
2 i \phi'_{V}(\sigma_1 + \sigma_2 - \sigma_1' - \sigma_2'),
\end{eqnarray}
where the primes on 
$G_{\sigma_1 \sigma_2 \sigma_1' \sigma_2'}(z;{\bf r}_1,{\bf
r}_2,{\bf r}_1',{\bf r}_2')$, $D(z;{\bf r})$ and $\phi_V$ denote derivatives
with respect to $z$ and the dependence of $\phi_V'$ on $z$ is implicit.

Changing co-ordinates to
\begin{eqnarray}
{\bf R} &=&{1 \over 4}({\bf r}_1 + {\bf r}_2 +{\bf r}_1' + {\bf r}_2'),
\nonumber \\
{\bf r} &=& {\bf r}_1 + {\bf r}_2 - {\bf r}_1' - {\bf r}_2', \nonumber \\
\brho_1 &=& {1 \over 2}({\bf r}_1 - {\bf r}_2 + {\bf r}_1' - {\bf r}_2'),
\nonumber \\
\brho_2 &=& {1 \over 2}({\bf r}_1 - {\bf r}_2 - {\bf r}_1' + {\bf r}_2'),
\end{eqnarray}
it is evident that $\Gamma_{\sigma_1 \sigma_2 \sigma_1' \sigma_2'}$ does
not depend on ${\bf R}$ in a homogeneous medium, and this also
enables us to eliminate ${\bf r}$ as
a parameter.  In view of this simplification it is convenient to
change notation, replacing $\gamma_{\sigma_1 \sigma_2 \sigma_1'
\sigma_2'}(z;{\bf R},{\bf r},
{\brho}_1,\brho_2)$ by $\Gamma_{\sigma_1 \sigma_2 \sigma_1' \sigma_2'}
(z;\brho_1,\brho_2)$.  Equation (\ref{4-2}) becomes
\begin{eqnarray}
\hfill \left[ 2 i k \frac{\partial}{\partial z} + 2 \, \nabla_{\brho_1}
\cdot \nabla_{\brho_2} + i k G'_{\sigma_1 \sigma_2
\sigma_1' \sigma_2'}(z;\brho_1,\brho_2) \right]
\Gamma_{\sigma_1 \sigma_2 \sigma_1' \sigma_2'}(z;{\brho}_1,\brho_2)  = 0, \hfill
\label{PropEqun} \label{4thcoher}
\end{eqnarray}
\begin{eqnarray}
G'_{\sigma_1 \sigma_2 \sigma_1' \sigma_2'}(z;\brho_1,\brho_2) &=&
D'_{\sigma_1 \sigma_1'}(z;\brho_2) + D'_{\sigma_1 \sigma_2'}(z;\brho_1)
+D'_{\sigma_2 \sigma_1'}(z;\brho_1) + D'_{\sigma_2
\sigma_2'}(z;{\brho}_2) \nonumber \\ &\null& -D'_{\sigma_1 \sigma_2}(
z;{\brho}_1+{\brho}_2) - D'_{\sigma_1' \sigma_2'}(z;\brho_1-\brho_2)+
2 i \phi'_{V}(\sigma_1 + \sigma_2 -\sigma_1' - \sigma_2').
\end{eqnarray}
The generalized fourth-order moment, $\Gamma_{\sigma_1 \sigma_2 \sigma_1'
\sigma_2'}
(z;\brho_1,\brho_2)$, describes the correlation
between four receivers arranged in a parallelogram whose axes
are $\brho_1$ and $\brho_2$ (see figure 1).  In an isotropic medium
the co-ordinates $\brho_1$ and
$\brho_2$ are interchangeable, $\Gamma_{\sigma_1 \sigma_2
\sigma_1' \sigma_2'}(z;\brho_1,{\brho}_2) = \Gamma_{\sigma_1 \sigma_2
\sigma_1' \sigma_2'}(z;\brho_2,\brho_1)$, (\S20-13 of \cite{Ishimaru}), but this
is not the case in general in an anisotropic medium.

It is useful to relate the generalized fourth-order moments to
auto- and cross-correlations of the Stokes visibilities.  The 16 terms
in $\Gamma_{\sigma_1 \sigma_2 \sigma_1' \sigma_2'}$ are separated
into four ($\sigma_1 = \sigma_1' = \sigma$, $\sigma_2 =
\sigma_2'=\sigma'$ with $\sigma=\pm 1, \sigma' = \pm 1$) that involve
only $I$ and $V$:
\begin{eqnarray}
\Gamma_{\sigma \sigma' \sigma \sigma'}(z,\brho_1,{\brho}_2) &=&
\langle I^2 \rangle (z,\brho_1,\brho_2)+
(\sigma+\sigma') \langle I V \rangle (z,{\brho}_1,\brho_2)
+ \sigma \sigma' \langle V^2 \rangle (z,\brho_1,\brho_2),
\label{StokesEqns1}
\end{eqnarray}
and four ($\sigma_1 = \sigma_2 =\sigma$, $\sigma_1' = \sigma_2' = \sigma'$
and $\sigma_1 = \sigma_2 = \sigma$, $\sigma_1'=\sigma_2'=\sigma'$
with $\sigma \neq \sigma'$) that involve only $Q$ and $U$:
\begin{eqnarray}
\Gamma_{\sigma \sigma \sigma' \sigma'}(z,\brho_1,{\brho}_2)
&=&
\langle Q^2 \rangle (z,\brho_1,\brho_2)
- 2 i \sigma \langle Q U \rangle (z,\brho_1,{\brho}_2) -
\langle U^2 \rangle (z,\brho_1,\brho_2),
\nonumber \\
\Gamma_{\sigma \sigma' \sigma' \sigma}(z,\brho_1,{\brho}_2)
&=&
\langle Q^2 \rangle (z,\brho_1,\brho_2)  +\langle U^2
\rangle (z,\brho_1,\brho_2).
\label{StokesEqns2}
\end{eqnarray}
The other eight terms involving cross correlations between $I$, $V$ and
$Q$, $U$ are not discussed here.

\subsection{Solution for a thin screen}
A standard approximation in scintillation theory
is to assume that the phase fluctuations in the medium
occur on a thin screen located a distance $z$ from the observer
\cite{Ishimaru,GoodNar}.
We assume that the incident wavefront is planar, corresponding to a
point source at $z=-\infty$ so that
$\langle XY \rangle (0,\brho_1,\brho_2)$ is
independent of the transverse spatial co-ordinates $\brho_1$ and
$\brho_2$.  This implies that the fourth-order moments incident on 
the screen are not functions of $\brho_1$ or ${\brho}_2$, and
we henceforth write $\langle XY \rangle (0)$ for the incident value.

If the source is located at a finite distance it is possible to
correct for the spherical nature of the wavefront by making the 
substitutions \cite{GoodNar}
\begin{eqnarray}
    z \rightarrow \frac{z_1 z_2}{z_1 + z_2}, \label{Corre1}\\
    {\bf r} \rightarrow \frac{z_1}{z_1 + z_2}{\bf r} \label{Corre2},
\end{eqnarray}
where $z_1$ is the distance from the source to the scattering screen and
$z_2$ the distance from the screen to the observer.  In particular,
equation (\ref{Corre2}) implies that the length scale of fluctuations on the
observer's screen is larger by a factor $(z_1+z_2)/z_1$ compared to the planar
case.

The solution of equation (\ref{PropEqun}) \cite{Ishimaru} for a
planar wavefront incident upon a thin screen located
a distance $z$ from the observer is
\begin{eqnarray}
\Gamma_{\sigma_1 \sigma_2 \sigma_1' \sigma_2'}(z,{\bf
r}_1,{\bf r}_2) &=&
\left( \frac{k}{2 \pi z}\right)^2 \int d^2 {\bf r}_1' d^2 {\bf r}_2'
\Gamma_{\sigma_1 \sigma_2 \sigma_1' \sigma_2'} (0)
\nonumber \\ &\null&
\qquad \qquad \times \exp \left[ \frac{ik({\bf r}_1-{\bf
r}_1')\cdot({\bf r}_2-{\bf r}_2')}{z} - \frac{1}{2} \int_0^z dz'
G'_{\sigma_1 \sigma_2 \sigma_1' \sigma_2'}(z;{\bf r}_1',{\bf r}_2') \right].
\end{eqnarray}
With this solution, we use equations
(\ref{StokesEqns1}) and (\ref{StokesEqns2}) to assemble solutions for the
correlations of
the Stokes visibilities as follows:
\begin{eqnarray}
\langle XY \rangle (z,{\bf r}_1,{\bf r}_2) &=& 
\left( \frac{k}{2 \pi z} \right)^2 
\int d^2{\bf r}_1' d^2{\bf r}_2' 
\exp \left[\frac{ik({\bf r}_1-{\bf r}_1')\cdot ({\bf r}_2 -
{\bf r}_2')}{z} - G_{II}({\bf r}_1',{\bf r}_2') \right] A_{XY}({\bf 
r}_1',{\bf r}_2'), \label{OnePolProp} \\
\left( \begin{array}{c} A_{II} \\ 
A_{IV} \\ A_{VV} 
\end{array} \right) &=&
\left( \begin{array}{ccc} 
W_{II} & 2 W_{IV} & W_{VV} \\
W_{IV} & W_{II}-W_{VV} & W_{IV} \\
W_{VV} & 2 W_{IV} & W_{II} \\ 
\end{array} \right)
\left( \begin{array}{c}
\langle I^2 \rangle (0) \\
\langle IV^2 \rangle(0) \\
\langle V^2 \rangle (0) \\
\end{array}
\right) \nonumber \\
\left( \begin{array}{c} A_{QQ} \\ 
A_{QU} \\ A_{UU} 
\end{array} \right) &=&
\left( \begin{array}{ccc} 
W_{QQ} & -2 W_{QU} & W_{UU} \\
W_{QU} & W_{QQ}-W_{UU} & -W_{QU} \\
W_{UU} & 2 W_{QU} & W_{QQ} \\ 
\end{array} \right)
\left( \begin{array}{c}
\langle Q^2 \rangle (0) \\
\langle QU^2 \rangle(0) \\
\langle U^2 \rangle (0) \\
\end{array}
\right)
\end{eqnarray}
where the arguments $({\bf r}_1',{\bf r}_2')$ are omitted, and we introduce
\begin{eqnarray}
\left( \begin{array}{c} W_{II} \\ W_{IV} \\ W_{VV} \end{array} \right) &=&
\frac{1}{2} \left( \begin{array}{c}
e^{-a} \cosh b + e^{a-c} \\
- e^{-a} \sinh b \\
e^{-a} \cosh b - e^{a-c} 
\end{array} \right) \nonumber \\
\left( \begin{array}{c} W_{QQ} \\ W_{QU} \\ W_{UU} \end{array} \right) &=&
\frac{e^a}{2} \left( \begin{array}{c}
e^{-c'} + d (f+f') \\
i \, d (f-f')/2  \\
e^{-c'} - d (f+f') 
\end{array} \right), \nonumber
\end{eqnarray}
\begin{eqnarray}
a = G_{VV}({\bf r}_1',{\bf r}_2'),  \qquad &\null&
b = 2 G_{IV}({\bf r}_1',{\bf r}_2'), \qquad
c = 2 D_{VV}({\bf r}_2'), \qquad
c' = 2D_{VV}({\bf r}_1'), \nonumber \\
d &=& D_{VV}({\bf r}_1'+{\bf r}_2') + D_{VV}({\bf r}_1'-{\bf r}_2')-
8C_{VV}(0), \nonumber \\
f &=& \exp[D_{IV}({\bf r}_1'+{\bf r}_2') - D_{IV}({\bf r}_1'-{\bf
r}_2')] e^{-4 i \phi_V}/2, \nonumber \\
f' &=& \exp[-D_{IV}({\bf r}_1'+{\bf r}_2') +
D_{IV}({\bf r}_1'-{\bf r}_2')] e^{4 i \phi_V}/2, \nonumber \\
\label{OneFreqWs}
\end{eqnarray}    
\begin{eqnarray}
G_{XY}({\bf r}_1,{\bf r}_2) &=& \frac{1}{2} \left[ 2 D_{XY}({\bf
r}_1) + 2 D_{XY}({\bf r}_2) - D_{XY}({\bf r}_1+{\bf r}_2) -
D_{XY}({\bf
r}_1-{\bf r}_2) \right], \qquad  X,Y=(I,V). \nonumber \\
\end{eqnarray}

In the absence of a magnetic field, an obvious, although important point is
that all the Stokes parameters
scintillate like the total intensity.  Referring back to the
definitions of $D_{IV}$ and $D_{VV}$ in \S3, the
absence of Faraday rotation terms implies that $D_{IV}$, $D_{VV}$
and $\phi_{V}$ are zero, leaving only $W_{II}$, $W_{QQ}$ and
$W_{UU}$ nonzero.  In particular, one has
$W_{II}=1$ and $W_{QQ}+W_{UU}=1$, so
the correlation functions differ only
by a multiplicative constant ($\langle X Y \rangle (0)$):
\begin{eqnarray}
\langle X Y \rangle (z,{\bf r}_1,{\bf r}_2)
 &=& \langle X Y \rangle (0) 
 \left( \frac{k}{2 \pi z} \right)^2 
\int d^2 {\bf r}_1' d^2 {\bf r}_2'
\exp \left[\frac{ik({\bf r}_1-{\bf r}_1')\cdot ({\bf r}_2 -
{\bf r}_2')}{z} - G_{II}({\bf r}_1',{\bf r}_2') \right],
\\
\hfill XY &=& [II,IV,VV,QQ,UU,QU]. \nonumber
\end{eqnarray}

\subsection{A simple model for rotation measure 
fluctuations}\label{CoherMag}
A simple model is when the fluctuations occur only in the electron
density, with the magnetic field being uniform.
In this case the structure functions $D_{VV}$ and $D_{IV}$ are related
to $D_{II}$ by the parameter $\alpha = \Omega_e/\omega$, where
$\Omega_e$ is the electron cyclotron frequency and $\omega$ is the
angular frequency of the radiation.  In particular, one has
$D_{VV}/\alpha^2 = D_{IV}/\alpha = D_{II}$.  Taking a typical value of the
magnetic field in the ISM of $3\,\mu$G and an observing frequency of
$\nu=1\,$GHz, $\alpha$ is of order $10^{-8}$.

In this case we expand equations (\ref{OneFreqWs})
in terms of $\alpha$, where $D_{IV}/D_{II}$ is of
order $\alpha$ and $D_{VV}/D_{II}$ is of order $\alpha^2$.  Retaining
terms to second order in $\alpha$ the second order correlations are 
given by equations (\ref{OnePolProp}) with 
\begin{eqnarray}
\left( \begin{array}{c} W_{II} \\ W_{IV} \\ W_{VV} \end{array} \right)     
&=& \left( \begin{array}{c} 
1 + b^2/4 - c/2 \\
b/2  \\ 
b^2/4 - a + c/2
\end{array} \right) \nonumber \\
\left( \begin{array}{c} W_{QQ} \\ W_{QU} \\ W_{UU} \end{array} \right)     
&=& \frac{1}{2} \left( \begin{array}{c} 
1 + a - c'  + g e^{-8C_{VV}(0)} \\
h e^{-8C_{VV}(0)} \\
1 + a - c' - g  e^{-8C_{VV}(0)}
\end{array} \right),
\label{Approxs} 
\end{eqnarray}
\begin{eqnarray}
g &=& \cos 4 \phi_{V} + i \sin 4 \phi_{V} \left(D_{IV}({\bf r}_1'-
{\bf r}_2')
-D_{IV}({\bf r}_1'+{\bf r}_2') \right)  \nonumber \\
&\null& \qquad \qquad  + \frac{1}{2} \cos 4 \phi_V
{\Big \{ } \left[ D_{IV}({\bf r}_1'-{\bf r}_2') - D_{IV}({\bf r}_1'+{\bf
r}_2')
\right]^2 \nonumber \\ &\null& 
\qquad \qquad \qquad -D_{VV}({\bf r}_1'+{\bf r}_2') -
D_{VV}({\bf r}_1' -{\bf r}_2') - G_{VV}({\bf r}_1',{\bf r}_2') 
{\Big \} } 
\nonumber \\
h &=& \sin 4 \phi_V - i \cos 4 \phi_V \left( D_{IV}({\bf r}_1'-{\bf r}_2')
-D_{IV}({\bf r}_1'+{\bf r}_2') \right)  \nonumber \\
&\null&  \qquad \qquad+ \frac{\sin 4 \phi_{V}}{2} {\Big \{ }
\left[ D_{IV}({\bf r}_1'-{\bf r}_2') - D_{IV}({\bf r}_1'+{\bf r}_2')
\right]^2  \nonumber \\ &\null&
\qquad \qquad \qquad + D_{VV}({\bf r}_1'+{\bf r}_2') +D_{VV}({\bf
r}_1'-{\bf r}_2') + G_{VV}({\bf r}_1',{\bf r}_2')  {\Big \} }.
\label{ApproxEnd}
\end{eqnarray}
Equations (\ref{Approxs}) and (\ref{ApproxEnd}) are used in the discussion below.

\section{Identification of Effects}
In this section we discuss the interpretation of the equations
(\ref{OnePolProp}) and (\ref{Approxs}), concentrating on the fluctuations
in circular polarization and the total intensity.

\subsection{Circular Polarization}
We are concerned with the creation of a circularly polarized component
and so we assume the source has no intrinsic circular polarization,
which corresponds to assuming $\langle
V^2 \rangle (0) = 0$ and $\langle IV \rangle (0) = 0$ in equation
(\ref{OnePolProp}).  The variances in the visibility and in the Stokes
$V$ visibility are then
\begin{eqnarray}
\langle I^2 \rangle(z,{\bf r}_1,{\bf r}_2) &=& 
\langle I^2 \rangle (0) \left( \frac{k}{2 \pi z} \right)^2 
\int d^2{\bf r}_1' {\bf r}_2' 
\exp \left[\frac{ik({\bf r}_1-{\bf r}_1')\cdot ({\bf r}_2 -
{\bf r}_2')}{z} - G_{II}({\bf r}_1',{\bf r}_2') \right] \nonumber \\
&\null& \qquad \qquad \times \left\{ 1 + \vert G_{IV}({\bf r}_1',{\bf
r}_2') \vert^2 - D_{VV}({\bf r}_2') \right\}, \label{Imain} \\
\langle V^2 \rangle(z,{\bf r}_1,{\bf r}_2) &=& 
\langle I^2 \rangle (0) \left( \frac{k}{2 \pi z} \right)^2 
 \int d^2{\bf r}_1' {\bf r}_2' 
\exp \left[\frac{ik({\bf r}_1-{\bf r}_1')\cdot ({\bf r}_2 -
{\bf r}_2')}{z} - G_{II}({\bf r}_1',{\bf r}_2') \right] \nonumber \\
&\null& \qquad \qquad \times \left\{ - G_{VV}({\bf r}_1,{\bf r}_2) +
D_{VV}({\bf r}_2') + \vert G_{IV}({\bf r}_1',{\bf r}_2') \vert^2  \right\}.
\label{Vmain}
\end{eqnarray}
For convenience in the discussion below, we label the contributions to
$\langle V^2 \rangle$
due to $G_{VV}$, $D_{VV}$ and $\vert G_{IV}\vert^2$
as $\langle V^2\rangle_1$, $\langle V^2 \rangle_2$ and $\langle V^2
\rangle_3$ in (\ref{Vmain}) respectively.

\subsubsection{Variations in the total intensity}
Before discussing the interpretation of equation (\ref{Vmain}), we
review the subject of intensity
variations in an isotropic medium because many of the
approximations and definitions are relevant to the
anisotropic case.  In the strong scattering regime, intensity
variations are due to
diffractive scintillation, caused by interference between subimages
over the scattering disk, and refractive scintillation, due to
refractive focussing and defocussing of the entire scattering disk
\cite{Narayan}.

Intensity variations in an isotropic medium correspond to the first term
in the curly brackets in
equation (\ref{Imain}).  Following \cite{TatZav} this term may be 
written in the form
\begin{eqnarray}
    \langle I^2 \rangle_{II} (z;{\bf r}_1,{\bf r}_2) &=& 
    \left( \frac{k}{2 \pi z} \right)^2 
    \int d^2 {\bf r}_1' d^2{\bf r}_2'
    \langle I^2 \rangle (0) 
    \exp \left[\frac{ik({\bf r}_1-{\bf r}_1')\cdot ({\bf r}_2 -
    {\bf r}_2')}{z} \right] \nonumber \\
    &\null& \qquad \qquad \times \left\{ e^{-D({\bf r}_1')}
    \left[1+\Omega({\bf r}_2',{\bf r}_1') + \cdots \right] +
    e^{-D({\bf r}_2')} \left[ 1 + \Omega({\bf r}_1',{\bf r}_2') +
    \cdots \right] \right\}, \label{Isqrapprox} \\
    \Omega({\bf r}_1,{\bf r}_2) &=& D_{II}({\bf r}_1+{\bf r}_2)/2 +
    D_{II}({\bf r}_1-{\bf r}_2)/2 - D_{II}({\bf r}_1), \label{OmegaDefn}
\end{eqnarray}
where the subscript $II$ signifies that only
isotropic phase fluctuations are retained.  The approximation made in
deriving equation (\ref{Isqrapprox}) follows by
recognizing that, for strong scattering, the intensity fluctuations
are dominated by regions where either ${\bf r}_1'$ or ${\bf r}_2'$
are small and then assuming $r_2 \gg r_1$ or $r_1 \gg r_2$ for these
regions respectively.
Equation (\ref{Isqrapprox}) is written compactly in the form
\begin{eqnarray}
  \langle I^2 \rangle_{\rm II}(z,{\bf r}_1,{\bf r}_2) =
  [2 + 2 \xi(z,{\bf r}_1,{\bf r}_2)]\langle I^2 \rangle (0).
  \label{Inormal}
\end{eqnarray}
The term $\xi$ is due to the contribution of $\Omega$, and
is associated
with large scale focussing and defocussing of the entire scattering
disk, giving rise to so-called ``refractive scintillation''
\cite{Narayan}.  The explicit expression for $\xi$ is
\begin{eqnarray}
\xi(z;{\bf r}_1,{\bf r}_2) =  2 \int
\frac{d^2 {\bf q}}{(2 \pi)^2} \Phi({\bf q}) \left\{ 1 - \cos({\bf q}
\cdot \brho_1 - q^2 r_{\rm F}^2)\right\} \exp[i {\bf q} \cdot
\brho_2 - D_{II}(\brho_1 - {\bf q}r_{\rm F}^2)].
\label{gammadefn}
\end{eqnarray}
This term is less than unity in the strong scattering
r\'egime \cite{Narayan}.
The power spectrum of phase inhomogeneities, $\Phi({\bf q})$, is
defined by
\begin{eqnarray}
D_{II}({\bf r}) = 2 \int \frac{d^2{\bf q}}{(2 \pi)^2} \Phi ({\bf q}) [1 - e^{i
{\bf q} \cdot {\bf r}} ]. \label{DIIPhidefn}
\end{eqnarray}
A power law spectrum $\Phi({\bf q})$ has the form
\begin{eqnarray}
\Phi ({\bf q}) &=& Q_{II} q^{-\beta}, \qquad \qquad q_{\rm min}<q<q_{\rm
max},
\label{Phidefn}
\end{eqnarray}
where $q_{\rm min}$ and $q_{\rm max}$ correspond to the inverses of
the outer and
inner scales of the scattering medium, $r_{\rm out}$ and $r_{\rm in}$,
respectively, and $Q_{II}$ is a constant.
The familiar case of Kolmogorov turbulence corresponds to $\beta=11/3$.
Relating the definition of the phase structure function $D_{II}({\bf r})$
provided by \cite{Colesetal} to the power spectrum of
phase inhomogeneities \cite{Goodmanetal}, one has
\begin{eqnarray}
Q_{II} = 8 \pi^3 \Delta L C_N^2 r_e^2 \lambda^2,
\end{eqnarray}
where $r_e$ is the classical radius of the electron, $\Delta L$ is the path
length through the scattering medium and $C_N^2$ is the electron
density structure constant.
The form (\ref{Phidefn}) for $\Phi({\bf q})$ is used extensively below.

Inspection of equation (\ref{Inormal}) reveals that the variance of the
intensity scintillations,
\begin{eqnarray}
\langle I^2 \rangle (z,0,0)- \langle I^2 \rangle (0) =
[1 + 2 \xi(z,0,0)]\langle I^2 \rangle(0)
\end{eqnarray}
contains contributions from two terms.
The term $2 \xi(z,0,0) \langle I^2 \rangle (0)$ is associated with
refractive scintillation, while the term of magnitude $\langle I^2 \rangle
(0)$ is identified with diffractive scintillation.

\subsubsection{Circular polarization correction terms}
We now consider the three terms in equation (\ref{Vmain}), assuming that
fluctuations
in the rotation measure are determined by fluctuations in the density
alone, implying $D_{II} = D_{IV}/\alpha = D_{VV}/\alpha^2$ with $\alpha =
\Omega_e/\omega$.

The contribution from the first term, denoted $\langle V^2
\rangle_1$, comes from the regions where
either ${\bf r}_1'$ or ${\bf r}_2'$ are small.  Since $G_{VV} =
\alpha^2 G_{II}$, this term may be
approximated by referring to equations (\ref{Isqrapprox}) and
(\ref{gammadefn}):
\begin{eqnarray}
\langle V^2 \rangle_1 (z,{\bf r}_1,{\bf r}_2) &=&  \alpha^2 
\left( \frac{k}{2 \pi z} \right)^2 
\int d{\bf r}_1' d {\bf r}_2' \langle I^2 \rangle (0)
\exp \left[\frac{ik({\bf r}_1-{\bf r}_1')\cdot ({\bf r}_2 -
{\bf r}_2')}{z} \right] \nonumber
\\
&\null& \qquad \times
\left\{ \exp[-D_{II}({\bf r}_1')] \Omega({\bf r}_2',{\bf r}_1')
+ \exp[-D_{II}({\bf r}_2')] \Omega({\bf r}_1',{\bf r}_2') \right\},
\label{V210simple}
\end{eqnarray}
which may be rewritten as
\begin{eqnarray}
\langle V^2 \rangle_1 (z,{\bf r}_1,{\bf r}_2) &=& \alpha^2 \langle I^2
\rangle (0) [\xi(z,{\bf r}_1,{\bf r}_2) + \xi(z,{\bf r}_2,{\bf
r}_1) ], \label{V21simple}
\end{eqnarray}
with $\xi$ given by equation (\ref{gammadefn}).  Production of
 circular polarization due to the term
$\langle V^2 \rangle_1$ is associated with the correction term in
equation (\ref{gammadefn}) due to refractive intensity
variations.

It is possible to derive an expression for $\langle V^2 \rangle_1$
from equation (\ref{V210simple}) directly.  Consider the
first term in curly brackets in equation
(\ref{V210simple}).  The exponential term is only large for small
${\bf r}_1'$, so it is appropriate to expand
$\Omega({\bf r}_2',{\bf r}_1')$ for small $r_1/r_2$:
\begin{eqnarray}
\Omega({\bf r}_2',{\bf r}_1') \approx \left( \frac{r_2'}{r_{\rm diff}}
 \right)^{\beta-2} (\beta-3) (\beta -2) \left( \frac{r_1'}{r_2'}
 \right)^2.
 \end{eqnarray}
Specializing to the case ${\bf r}_1 = {\bf r}_2 = 0$, the terms
involving
${\bf r}_1'$ and ${\bf r}_2'$ in curly brackets in equation (\ref{V210full})
are interchangeable.  One then has
\begin{eqnarray}
\langle V^2 \rangle_1 (z,0,0) &=& 2 \alpha^2 \langle I^2 \rangle (0)
\left( \frac{k}{2 \pi z} \right)^2 \int d{\bf r}_1'
(\beta -3) (\beta-2) r_1^2  \nonumber \\
&\null& \times \left( \frac{1}{r_{\rm diff}}\right)^{\beta-2}
\int d {\bf r}_2' \exp \left[\frac{ik{\bf r}_1'\cdot {\bf r}_2'}{z} \right]
{\bf r}_2'^{\beta-4}.
\end{eqnarray}
We evaluate the integral over ${\bf r}_2'$ using
\cite{Lighthill}
\begin{eqnarray}
\int d^2{\bf r}_2' \exp \left[ i {\bf x} \cdot {\bf r}_2'
\right] r_2^{\gamma} =  \pi 2^{\gamma+2}
\frac{\Gamma(\gamma/2+1)}{\Gamma(-\gamma/2)} x^{-\gamma-2}, 
\label{Lightequn}
\end{eqnarray}
which is valid for $\gamma > -2$.  The remaining integral
over ${\bf r}_1'$ yields
\begin{eqnarray}
\langle V^2 \rangle_1 (z,0,0) &=& \alpha^2
\langle I^2 \rangle (0) 2^{\beta-3} \pi^{-1} (\beta-3)
(\beta-2) \left( \frac{r_{\rm F}}{r_{\rm diff}} \right)^{2 \beta - 8}
\frac{ \Gamma(\beta/2 -1)}{
\Gamma(2-\beta/2)} \Gamma \left( \frac{6 -\beta}{\beta-2}\right).
\label{FinV21}
\end{eqnarray}
Since the modulation index due to refractive intensity
variations, $m_{\rm ref}$, is of order $(r_{\rm F}/r_{\rm diff})^{4-\beta}$
(see, e.g., \cite{GoodNar}), equation (\ref{FinV21}) implies
a degree of circular polarization of order $\alpha m_{\rm ref}
\langle I^2 \rangle^{1/2}$.  Appendix A details the contribution
of $\langle V^2 \rangle_1$ when the
rotation measure fluctuations are not necessarily proportional to
the isotropic phase fluctuations.

The contribution due to the term $\langle V^2 \rangle_2$ is
also evaluated by representing the rotation measure structure
function in terms of its power spectrum.  Evaluating the integrals
over ${\bf r}_1'$ and ${\bf r}_2'$ in equation (\ref{Vmain}), the variance
in circular polarization due to the term $D_{VV}({\bf r}_2')$ is
\begin{eqnarray}
\langle V^2 \rangle_{2} (z,{\bf r}_1,{\bf r}_2) &=& 2 \alpha^2 \langle
I^2 \rangle (0) \int d^2 {\bf q} \frac{\Phi({\bf q})}{(2 \pi)^2}
{\bigg [}
\exp[-D({\bf r}_1)] \nonumber \\
&\null&  \qquad -
\frac{1}{2} \left\{ e^{i {\bf q}\cdot {\bf r}_2} \exp[-D({\bf r}_1-
r_{\rm F}^2 {\bf q}) + e^{-i {\bf q}\cdot {\bf r}_2} \exp[-D({\bf
r}_1+r_{\rm F}^2 {\bf q}) \right\} {\bigg ]}. \label{interimV2}
\end{eqnarray}
For arbitrary ${\bf r}_1$ and ${\bf r}_2$ this expression is, in
general, complex, and unlike $\langle V^2 \rangle_1(z,{\bf r}_1,{\bf
r}_2)$, it is not symmetric under interchange of
${\bf r}_1$ and ${\bf r}_2$.  The circular polarization is a real
quantity, and equation (\ref{interimV2}) can be directly
expressed in terms of the Stokes parameter $V$ only for ${\bf r}_2=0$,
in which case the expression is real, as required.

The contribution from $\langle V^2 \rangle_2$ to the variance in circular
polarization is obtained by setting
${\bf r}_1={\bf r}_2=0$ in equation (\ref{interimV2}).  The integral
is approximated using
\begin{eqnarray}
\left[ 1 - \exp[-D({\bf q} r_{\rm F}^2)] \right] &\approx& \left\{
\begin{array}{ll}
D({\bf q} r_{\rm F}^2), & q_{\rm min}<q < 1/r_{\rm ref}, \\
1, & 1/r_{\rm ref} < q <q_{\rm max},
\end{array}
\right.
\end{eqnarray}
to yield
\begin{eqnarray}
\langle V^2 \rangle_2 (z,0,0) &=& \frac{\alpha^2
\langle I^2
\rangle (0)}{\pi}
(r_{\rm ref})^{\beta-2} \left[\frac{1}{\beta -2}  + \ln \left(
\frac{1}{q_{\rm min} r_{\rm ref}}\right)
\right].
\end{eqnarray}
where  we assume that the outer scale is much larger
than the refractive scale ($L_0 \gg r_{\rm ref}$).  Rewriting 
$Q_{II}$ in terms of
$r_{\rm diff}$, one finds the magnitude of the circular
polarization:
\begin{eqnarray}
\langle V^2 \rangle_2 (z,0,0) &=& \alpha^2  \langle I^2 \rangle
(0) f(\beta) \left(\frac{r_{\rm F}}{r_{\rm diff}} \right)^{2\beta-4}
\left[ \frac{1}{\beta-2} + \ln
\left( \frac{1}{q_{\rm min} r_{\rm ref}} \right) \right],
\label{V2result}
\end{eqnarray}
with
\begin{eqnarray}
f(\beta) = - 2^{\beta-1} \frac{\Gamma(\beta/2)}{\Gamma(1-\beta/2)},
\end{eqnarray}
for $\beta<4$.  For $\beta=11/3$ one has $f(\beta) \approx 0.89$.
The contribution of the circular polarization due to this term scales
as $\alpha^2 (r_{\rm ref}/r_{\rm diff})^{\beta-2}$.

The contribution of $\langle V^2 \rangle_3$ in a homogeneous magnetic
field is evaluated by appealing to the arguments presented in the
evaluation of $\langle V^2 \rangle_1$.  For strong scattering, only
the regions near ${\bf r}_1' \approx 0$ and ${\bf r}_2' \approx 0$
contribute.  Expanding for $r_1' \ll r_2'$ and $r_2' \ll r_1'$ one has
\begin{eqnarray}
\langle V^2 \rangle_3(z,{\bf r}_1,{\bf r}_2) &=& \alpha^2
  \langle I^2 \rangle (0) 
  \left( \frac{k}{2 \pi z} \right)^2 \int d^2 {\bf r}_1'
d^2{\bf r}_2'
    \exp \left[\frac{ik({\bf r}_1-{\bf r}_1')\cdot ({\bf r}_2 -
{\bf r}_2')}{z} \right] \nonumber \\
    &\null& \qquad \qquad \times \left\{ e^{-D({\bf r}_1')}
    \Omega^2({\bf r}_2',{\bf r}_1') +
    e^{-D({\bf r}_2')} \Omega^2({\bf r}_1',{\bf r}_2') \right\}.
    \label{V23gen}
\end{eqnarray}
Consider the second term in curly brackets in (\ref{V23gen}).  The
exponential ensures that this term is only large for small $r_2'$,
so one may expand $\Omega^2 ({\bf r}_1',{\bf r}_2')$, defined by
equation (\ref{OmegaDefn}), for small $r_2'$.  To
second order in $r_2'/r_1'$ one has
\begin{eqnarray}
\Omega^2({\bf r}_1',{\bf r}_2') &\approx&
\left( \frac{r_1'}{r_{\rm diff}} \right)^{2 (\beta-2)}
\frac{(\beta-2)^2 (\beta-3)^2 r_2'^4}{4 r_1'^4}.
\end{eqnarray}
Recognizing that the two terms inside the curly brackets in
equation (\ref{V23gen}) are identical
under the replacement ${\bf r}_1' \leftrightarrow {\bf r}_2'$, provided
${\bf r}_1 = {\bf r}_2=0$, one then has
\begin{eqnarray}
\langle V^2 \rangle_{3}(z,0,0) &\approx&
2 \left( \frac{k }{2 \pi z}\right)^2 \alpha^2 \int
d^2 {\bf r}_1' d^2 {\bf r}_2'
\exp \left[ \frac{ik{\bf r}_1' \cdot {\bf r}_2'}{z} - D_{II}({\bf r}_1) \right]
 \frac{r_1^4 D_{II}^2({\bf r}_2)}{r_2^4} \frac{(\beta-2)^2 
 (\beta-3)^2}{4}.
\end{eqnarray}
Now we may evaluate the integral over ${\bf r}_2'$ using equation 
(\ref{Lightequn}).  Then equation (\ref{V23gen}) reduces to
\begin{eqnarray}
\langle V^2 \rangle_3 = \pi \langle I^2 \rangle (0)
\frac{\alpha^2 (\beta-2)^2 (\beta-3)^2}{8 \pi^2}
r_{\rm F}^{4 \beta -16}
r_{\rm diff}^{4 - 2 \beta}  2^{2 \beta-6}
\frac{\Gamma(\beta-3)}{\Gamma(4-\beta)}
\int d^2{\bf r}_1 r_1^{10-2 \beta} \exp[-D_{II}({\bf r}_1')].
\end{eqnarray}
Now, using
\begin{eqnarray}
\int_0^{\infty} dr_1 \, r_1^{\gamma} \exp[- (r/r_{\rm diff})^{\beta-2}] =
r_{\rm diff}^{1 + \gamma} \frac{1}{\beta-2} \Gamma \left(
\frac{1+\gamma}{\beta-2} \right),
\end{eqnarray}
which is valid for $\beta>2$, one has
\begin{eqnarray}
\langle V^2 \rangle_3&=&  \alpha^2 \langle I^2 \rangle (0) (\beta-2)
(\beta-3)^2 2^{2 \beta -7} \left(
\frac{r_{\rm F} }{r_{\rm diff}} \right)^{4 \beta-16}  \Gamma\left( \frac{12
- 2\beta}{\beta-2}\right)
\frac{\Gamma(\beta-3)}{\Gamma(4-\beta)}. \label{V23res}
\end{eqnarray}
For strong scattering, the term
$(r_{\rm F} / r_{\rm diff})^{4 \beta-16}$ is less than unity.  In
particular, for a Kolmogorov spectrum of turbulent fluctuations
($\beta=11/3$) the root mean square degree of circular polarization due to
this term
scales as $\alpha (r_{\rm diff}/r_{\rm F})^{2/3}$, less than unity
for strong scattering.

The dominant term is $\langle V^2 \rangle_2 \gg \langle V^2 \rangle_1, 
\langle V^2 \rangle_3$ as this is the only term that allows 
$\sqrt{\langle V^2 \rangle} / \langle I \rangle \alpha$ to be much 
greater than unity.  We retain only this dominant term in the 
discussion below.

\subsubsection{Cross Correlation}

So far we discuss only the variance $\langle V^2 \rangle$ in Stokes
$V$.  The correlation function $\langle I V \rangle$ contains
additional information about the propagation-induced circular polarization.

The form of the expression for $\langle I V \rangle (z,{\bf r}_1,{\bf
r}_2)$ in equation (\ref{OnePolProp}), in
the absence of any intrinsic circular polarization, is
of the same form as the expression for $\langle V^2 \rangle_1$, yielding (cf.
equation (\ref{V210simple}))
\begin{eqnarray}
\langle I V \rangle (z,0,0) &=&  
\langle I^2 \rangle (z,0,0) \left( \frac{k}{2 \pi z} \right)^2 
\int d^2 {\bf r}_1' d^2{\bf r}_2' \, 
\exp \left[ \frac{ik{\bf r}_1' \cdot {\bf r}_2'}{z} \right] \nonumber \\
&\null& \qquad \qquad \times \left\{ \exp[-D({\bf r}_1')]
\Omega_{IV}({\bf r}_2',{\bf r}_1') + \exp[-D({\bf r}_2')]
\Omega_{IV}({\bf r}_1',{\bf r}_2')\right\}, \\
\Omega_{IV}({\bf r}_1,{\bf r}_2) &=& D_{IV}({\bf r}_1+{\bf r}_2) +
D_{IV}({\bf r}_1 - {\bf r}_2) - D_{IV}({\bf r}_1).
\end{eqnarray}
For a homogeneous magnetic field, one has $\Omega_{IV}({\bf
r}_1,{\bf r}_2) = \alpha
\Omega({\bf r}_1,{\bf r}_2)$ and the cross correlation is
\begin{eqnarray}
\langle I V \rangle (z,0,0) &=&  2 \alpha \xi(z,0,0) \langle I^2
\rangle(0). \label{IVcorres}
\end{eqnarray}
Since $\xi(z,0,0)$ is of order unity, it follows that the cross 
correlation is of order $\alpha$, rendering this correlation too small 
to be observed.

\section{Discussion and Conclusions}
Scintillations in a magnetized ISM necessarily lead to a small degree
of circular polarization (CP) even for an unpolarized source.  This occurs
because the refractive indices for the two natural wave modes are
slightly different, leading to a variety of small effects that are
different for the two opposite CP components.
Under the assumption that the fluctuations involve only the plasma density,
these effects are characterized by a single small parameter,
$\alpha=\Omega_e/\omega$, which is the ratio of the electron cyclotron
frequency at the scattering screen to the wave frequency.  For typical
parameters in the ISM, $\alpha \sim 10^{-8}$ is too small for the
induced CP to be of practical interest unless the
effect can be enhanced in some way.

The effect that we identify as of possible practical interest appears
in the variance of the Stokes parameter $V$, denoted $\langle V^2
\rangle$.  Provided that the time scale of observation is short compared
with the time scale on which $\langle V^2 \rangle$ changes, a net circular
polarization of order $\langle V^2 \rangle^{1/2}/\langle I \rangle$ should
be observed.  In evaluating $\langle V^2 \rangle$ we separate it into
three terms,
$\langle V^2 \rangle_i, i=1,2,3$, and find that each of them is of the
form $\langle V^2 \rangle_i = \alpha^2 \langle I^2 \rangle (0) A_i
(r_{\rm F}/r_{\rm diff})^{a_i}$, with the $A_i$ all of order unity.
For strong scattering, $r_{\rm F} \gg r_{\rm diff}$ implies 
$\langle V^2 \rangle \gg \alpha^2 \langle I^2 \rangle$
provided that $a_i$ is not too small.  For the term $\langle V^2
\rangle_2$, with $a_2=2\beta -4=10/3$ for a Kolmogorov spectrum
$\beta=11/3$, the numerical factor is large when the scattering
is strong.  This term arises from scattering due to the largest
structures in the postulated power-law spectrum of fluctuations.

An obvious prediction of this theory is that the observed
degree of CP should reverse randomly on the time
scale over which $\langle V^2 \rangle$ changes. However, the simplifying
assumptions we make need to be reconsidered in making a realistic estimate
of both the magnitude and time scale of the fluctuations in $V$. In
particular, we assume that both the fluctuations that lead to
scintillations in the usual sense and the fluctuations in rotation measure
that cause a separation in the right and left polarizations occur at a
single phase screen. The assumption that these two effects occur together
is unnecessarily restrictive. The effect that we describe requires
(a)~that the wavefront be rippled, and (b)~that the right and left
polarizations be separated, but there is no need for these two effects to
be due to the same structures in the ISM. Indeed, the dominant
contribution to $\langle V^2 \rangle$ is due to the large-scale
structures in the assumed turbulence, and in fact any large-scale
magnetized structure can act as a ``Faraday wedge'' in refracting the right
and left polarizations into slightly different directions. In principle,
the separation between the two circularly polarized rays can become
arbitrarily large at arbitrarily large distances from such a Faraday
wedge, leading to non-overlapping right and left hand polarized images,
and so to $\langle V^2 \rangle\sim\langle(I-\langle I\rangle)^2 \rangle$.
We propose to develop this idea further and apply it to the interpretation
of observed CP elsewhere.

Furthermore, an exact theory requires knowledge of the turbulent fluctuations 
in both the orientation and magnitude of the magnetic field 
(here assumed to be homogeneous), and its correlation with density 
fluctuations.	The turbulence in the ISM is expected to be highly anisotropic 
with respect to the magnetic field lines.  As remarked in \cite{GoldShrid}, the 
fact that the VLBI images of scatter-broadened images are typically only 
slightly elliptical (with axial ratios at most $\sim$~2:1) implies
that the radiation propagates through many regions with different 
magnetic field orientations.  This is expected to enhance  
the production of scintillation-induced CP.

We conclude (1)~that scintillation-induced CP must occur
due to propagation through the magnetized ISM, (2)~that the simplest
estimate that it is of order $\alpha\sim10^{-8}$ and so would be
unobservably small, is correct only for the mean value of $\langle
V\rangle/\langle I\rangle$, (3)~that the variance $\langle
V^2\rangle/\langle(I-\langle I\rangle)^2\rangle$ can be much larger than
$\alpha^2$, and (4)~that a more general model that treats the
scintillations and the separation of the opposite circularly polarized
rays (``Faraday wedge'') in different ways is needed in formulating a theory
that will be a useful basis for comparison with observations.

\acknowledgments
We thank Mark Walker for the suggesting the investigating the effect of rotation
measure gradients and Ron Ekers for helpful discussions.

\appendix

\section{The contribution of $\langle V^2 \rangle_1$}
We evaluate the contribution of the $\langle V^2 \rangle_1$
term when the structure functions $D_{II}$ and $D_{VV}$ are not necessarily
proportional to each other.  Writing
\begin{eqnarray}
    D_{VV}({\bf r}) &=& \left( \frac{r}{r_{\rm diffR}} \right)^{\beta_V - 2},
\end{eqnarray}
the contribution of $\langle V^2 \rangle_1$ is
\begin{eqnarray}
\langle V^2 \rangle_1 (z,{\bf r}_1,{\bf r}_2) &=& 
\int d{\bf r}_1' d {\bf r}_2' \langle I^2 \rangle (0)
\exp \left[\frac{ik({\bf r}_1-{\bf r}_1')\cdot ({\bf r}_2 -
{\bf r}_2')}{z} \right] \\
&\null& \qquad \times
\left\{ \exp[-D_{II}({\bf r}_1')] \Omega_{VV}({\bf r}_2',{\bf r}_1')
+ \exp[-D_{II}({\bf r}_2')] \Omega_{VV}({\bf r}_1',{\bf r}_2') \right\},
\label{V210full}
\end{eqnarray}
where we write
\begin{eqnarray}
\Omega_{VV}({\bf r}_2,{\bf r}_1) &=& D_{VV} ({\bf r}_1 + {\bf r}_2)/2 +
D_{VV}({\bf r}_1 - {\bf r}_2)/2 - D_{VV}({\bf r}_2).
\end{eqnarray}

Consider the first term in curly brackets in equation
(\ref{V210full}).  The exponential term is only large for small ${\bf r}_1$, so
it is appropriate to expand $\Omega_{VV}({\bf r}_2,{\bf r}_1)$ for
small $r_1/r_2$:
\begin{eqnarray}
\Omega_{VV}({\bf r}_2,{\bf r}_1) \approx \left( \frac{r_2}{r_{\rm diffR}}
 \right)^{\beta_V-2} (\beta_V-3) (\beta_V -2) \left( \frac{r_1}{r_2}
 \right)^2.
\end{eqnarray}
Specializing to the case ${\bf r}_1 = {\bf r}_2 = 0$, the terms
involving
${\bf r}_1'$ and ${\bf r}_2'$ in curly brackets in equation (\ref{V210full})
are interchangeable.  One then has
\begin{eqnarray}
\langle V^2 \rangle_1 (z,0,0) &=& 2 \langle I^2 \rangle (0)
\left( \frac{k}{2 \pi z} \right)^2 \int d{\bf r}_1'
(\beta_V -3) (\beta_V-2) r_1^2  \nonumber \\
&\null& \left( \frac{1}{r_{\rm diffR}}\right)^{\beta_V-2}
\int d {\bf r}_2' \exp \left[\frac{ik{\bf r}_1'\cdot {\bf r}_2'}{z} \right]
{\bf r}_2'^{\beta_V-4}.
\end{eqnarray}
We evaluate the integral over ${\bf r}_2'$ using (\ref{Lightequn}).  The 
remaining integral over ${\bf r}_1'$ is evaluated to yield
\begin{eqnarray}
\langle V^2 \rangle_1 (z,0,0) &=&
\langle I^2 \rangle (0) 2^{\beta_V-3} \pi^{-1} (\beta_V-3)
(\beta_V-2) r_{\rm F}^{2 \beta_V - 8} r_{\rm diffR}^{2 -\beta_V}
r_{\rm diff}^{6-\beta_V} \frac{ \Gamma(\beta_V/2 -1)}{
\Gamma(2-\beta_V/2)} \Gamma \left( \frac{6
-\beta_V}{\beta-2}\right). \label{FinV21app}
\end{eqnarray}
Setting $D_{VV}({\bf r}) = \alpha^2 D_{II}({\bf r})$, one then has
$\beta_V = \beta$ and $r_{\rm diffR}^{2-\beta} =
\alpha^2 r_{\rm diffI}^{2-\beta}$, and equation (\ref{FinV21app}) is then
proportional to the variance due to refractive scintillation, of
order $(r_{\rm F}/r_{\rm diff})^{8-2\beta}$ which is consistent with
equation (\ref{FinV21}).

\begin{figure}
\psfig{file=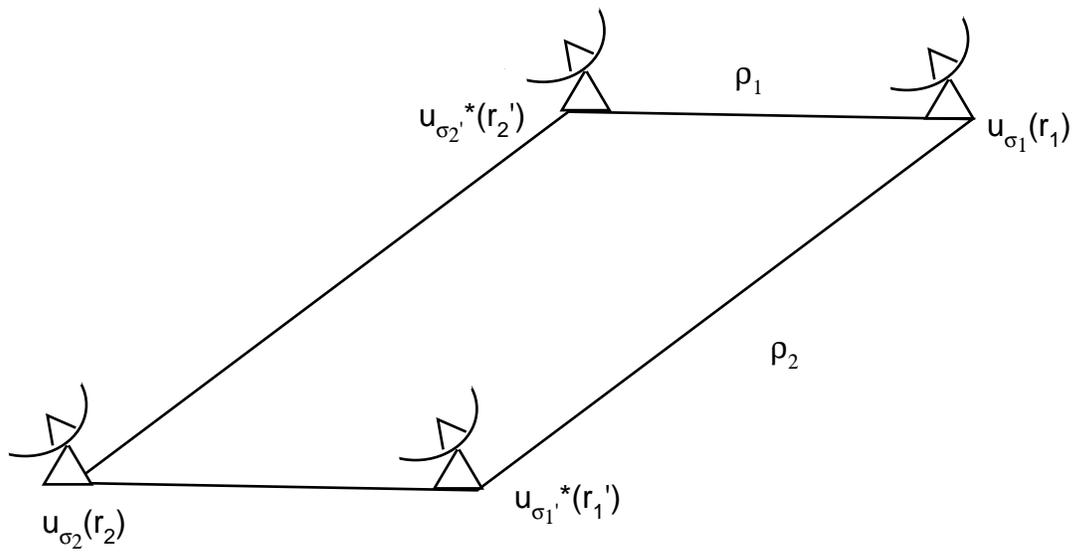}
\caption{Positions of the receivers in calculating
fourth-order moments (cf. Ishimaru 1978).}
\end{figure}

\end{document}